\documentclass[letterpaper, 10 pt, conference]{ieeeconf}  

\IEEEoverridecommandlockouts                              

\usepackage{cite}
\usepackage{amsmath,amssymb,amsfonts}
\usepackage{algorithmic}
\usepackage{graphicx}
\usepackage{textcomp}
\usepackage{xcolor}
\usepackage{comment}

\usepackage{caption}
\usepackage{subcaption}
\usepackage{multirow} 
\usepackage{booktabs} 
\usepackage{breqn}
\usepackage{csquotes}

\usepackage{hyperref}
\usepackage{cleveref}

\title{\LARGE \bf
Reasoning LLMs for User-Aware Multimodal Conversational Agents
}

\author{Hamed Rahimi$^{1,^*}$, Jeanne Cattoni$^{2,3}$, Meriem Beghili$^{1}$, Mouad Abrini$^{1}$, \\
Mahdi Khoramshahi$^{1}$, Maribel Pino$^{3}$, and Mohamed Chetouani$^{1}$
\thanks{$^*$Correspondence: {\tt\footnotesize hamed.rahimi@sorbonne-universite.fr}}
\thanks{$^{1}$ISIR, Sorbonne Université, Paris, France.} 
\thanks{$^{2}$Université Paris Cité, Paris, France.}
\thanks{$^{3}$Assistance Publique – Hôpitaux de Paris (AP-HP), Paris, France.
}}

\begin{document}

\maketitle


\begin{abstract}

Personalization in social robotics is critical for fostering effective human-robot interactions, yet systems often face the \textit{cold start} problem, where initial user preferences or characteristics are unavailable. This paper proposes a novel framework called USER-LLM R1 for a user-aware conversational agent that addresses this challenge through dynamic user profiling and model initiation. Our approach integrates chain-of-thought (CoT) reasoning models to iteratively infer user preferences and vision-language models (VLMs) to initialize user profiles from multimodal inputs, enabling personalized interactions from the first encounter. Leveraging a Retrieval-Augmented Generation (RAG) architecture, the system dynamically refines user representations within an inherent CoT process, ensuring contextually relevant and adaptive responses. Evaluations on the \textit{ElderlyTech-VQA Bench} demonstrate significant improvements in ROUGE-1 (+23.2\%) ROUGE-2 (+0.6\%) and ROUGE-L (+8\%) F1 scores over state-of-the-art baselines, with ablation studies underscoring the impact of reasoning model size on performance. Human evaluations further validate the framework’s efficacy, particularly for elderly users, where tailored responses enhance engagement and trust. Ethical considerations, including privacy preservation and bias mitigation, are rigorously discussed and addressed to ensure responsible deployment. 
\end{abstract}


\section{INTRODUCTION}
The integration of personalized and adaptive AI systems in human-robot interactions (HRI) is pivotal for fostering ethical, safe, and contextually effective engagements~\cite{irfan2019personalization,lee2012personalization}. Personalization necessitates that robots dynamically adjust their behavior to align with individual user characteristics, such as age, cognitive abilities, and cultural background, ensuring interactions are both socially appropriate and functionally optimized~\cite{rahimi2025user}. For instance, a child-centric interaction may demand simplified language, playful tones, and heightened safety protocols, whereas adult interactions could prioritize task complexity, formal dialogue, and nuanced social cues. Achieving such adaptability requires mechanisms to tailor actions while addressing challenges like data privacy, algorithmic bias, and transparency in decision-making~\cite{rahimi2025userAAAI}. User modeling aims to discover patterns and learn representations from user data, capturing characteristics such as profiles, preferences, and personalities~\cite{fischer2001user}.

Recent advancements in Multimodal Large Language Models (MLLMs)~\cite{caffagni2024revolution} have significantly enhanced the capacity of robots to interpret textual and visual cues, enabling more intuitive and adaptive interactions~\cite{alaluf2024myvlm,ning2024user}. However, critical gaps persist when user modeling metadata (e.g. demographic traits, behavioral patterns, or interaction histories) are absent, creating a "cold-start" problem~\cite{wongchokprasitti2015user} that limits personalization efficacy. Even when partial information exists, current systems often rely excessively on prior conversational data, neglecting dynamic methods to infer, validate, or update user profiles in real time~\cite{wang2022attention}. In \cite{rahimi2025user}, a set of Vision Language Models (VLMs) are trained to provide personalized responses corresponding to implicit user modeling of age, gender, emotion, and ethnicity trough user images. Despite bias optimization of these VLMs, a seamless interaction requires a dynamic mechanism that builds and confirms user profiles before inference while engaging with the user. 
\begin{figure}
    \centering
    \includegraphics[width=1\linewidth]{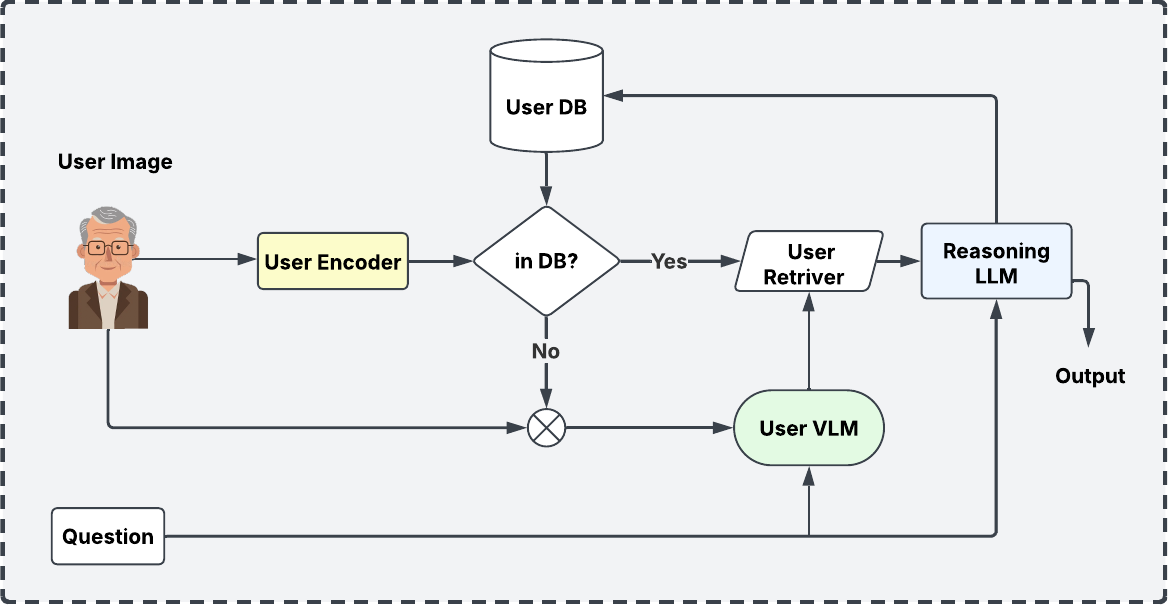}
    \caption{\textbf{USER-LLM R1 Architecture:} The framework consists of three principal components: a User Encoder for profile encoding, a Vision-Language Model (VLM) for initial user modeling, and a Chain-of-Thought (CoT) Reasoning Large Language Model (LLM) for updating user profile and personalized response generation.  }
    \label{fig:arch}
\end{figure}
As shown in \Cref{fig:arch}, our proposed framework addresses the complex challenge of personalized human-robot interaction through a novel multi-stage reasoning approach. This mechanism follows a deliberate chain of steps that tracks different users, infers initial user models, validates or refines these models through interaction, and ultimately delivers personalized responses. Central to our approach is a four-component architecture: (1) a user encoder that effectively differentiates between user profiles; (2) a RAG (retrieval augmented generation) process that infers user profiles and leverages previous conversations; (3) a User decoder (USER-VLM) that generates initial user models to address the "cold start" problem; and (4) a Chain of Thought reasoning process that holistically considers all aspects of user profiles to provide personalized responses. This framework specifically distinguishes between priors and posteriori intuitions about users, enabling robots to adapt their responses cautiously, for instance, by verifying assumptions, while progressively refining their user models. To overcome the dual challenges of personalization and privacy, our system autonomously gathers and synthesizes user insights during interactions in an implicit or explicit manner.

\section{RELATED WORK}

Recent advances in Large Language Models (LLMs) have significantly impacted user modeling and personalization approaches~\cite{tan2023user}. This section reviews the relevant literature on LLM-based user modeling techniques, with emphasis on RAG systems, model personalization frameworks, and reasoning-enhanced approaches.

\subsection{User Modeling with Large Language Models}

User Models enable personalization in various applications including recommendation systems, education, and healthcare. With the emergence of foundation models demonstrating superior performance in generating, understanding, and reasoning over multimodal data, user modeling approaches have been enhanced through integration with LLMs~\cite{tan2023user}. Current approaches to LLM-based user modeling can be categorized into two main streams. The first approach incorporates user history and preferences directly into prompts or uses retrieval systems to enhance contextual understanding \cite{comendant2024large,zerhoudi2024personarag}. The second approach tackles personalization from a model-centric perspective by developing specialized user encoders or modifying existing model architectures to inherently provide personalized responses based on user features, whether textual or visual~\cite{alaluf2024myvlm,rahimi2025user}.


Retrieval-augmented generation (RAG) has emerged as a powerful technique for enhancing the accuracy and reliability of generative AI models by incorporating information from specific and relevant data sources~\cite{arslan2024survey}. Several frameworks have demonstrated the effectiveness of RAG in personalization contexts. PersonaRAG~\cite{zerhoudi2024personarag} introduces user-centric agents that adapt both retrieval and generation processes based on real-time user data and interactions. Similarly, the "GenUI(ne) CRS" framework~\cite{maes2024genui} leverages LLMs for adaptive and interactive user interfaces, supporting domain-specific graphical elements while addressing the knowledge cut-off problem through RAG implementation. In domain-specific applications, such as swimming coaching systems, RAG has been utilized to capture user preferences and behaviors, significantly improving the quality of LLM outputs by leveraging contextual and real-world knowledge~\cite{comendant2024large}. These applications demonstrate RAG's versatility in enhancing personalization across diverse use cases.

Recent advancements in personalized LLMs emphasize architecture design and training tailored to individual user preferences and contextual histories. HYDRA~\cite{zhuang2024hydra} employs model factorization to combine user-specific behavioral patterns with general knowledge, using a reranker to prioritize relevant historical data and an adapter to align outputs with user preferences. User-LLM~\cite{ning2024user} introduces user embeddings derived from self-supervised learning to capture latent user behaviors and dynamically contextualize LLMs via cross-attention mechanisms. For multimodal interactions, User-VLM 360°~\cite{rahimi2025user} integrates user modeling with visual-linguistic signals, bias-aware tuning, and socio-emotive datasets to optimize real-time, contextually aware, and equitable interactions.

\begin{figure}
    \centering
    \includegraphics[width=1\linewidth]{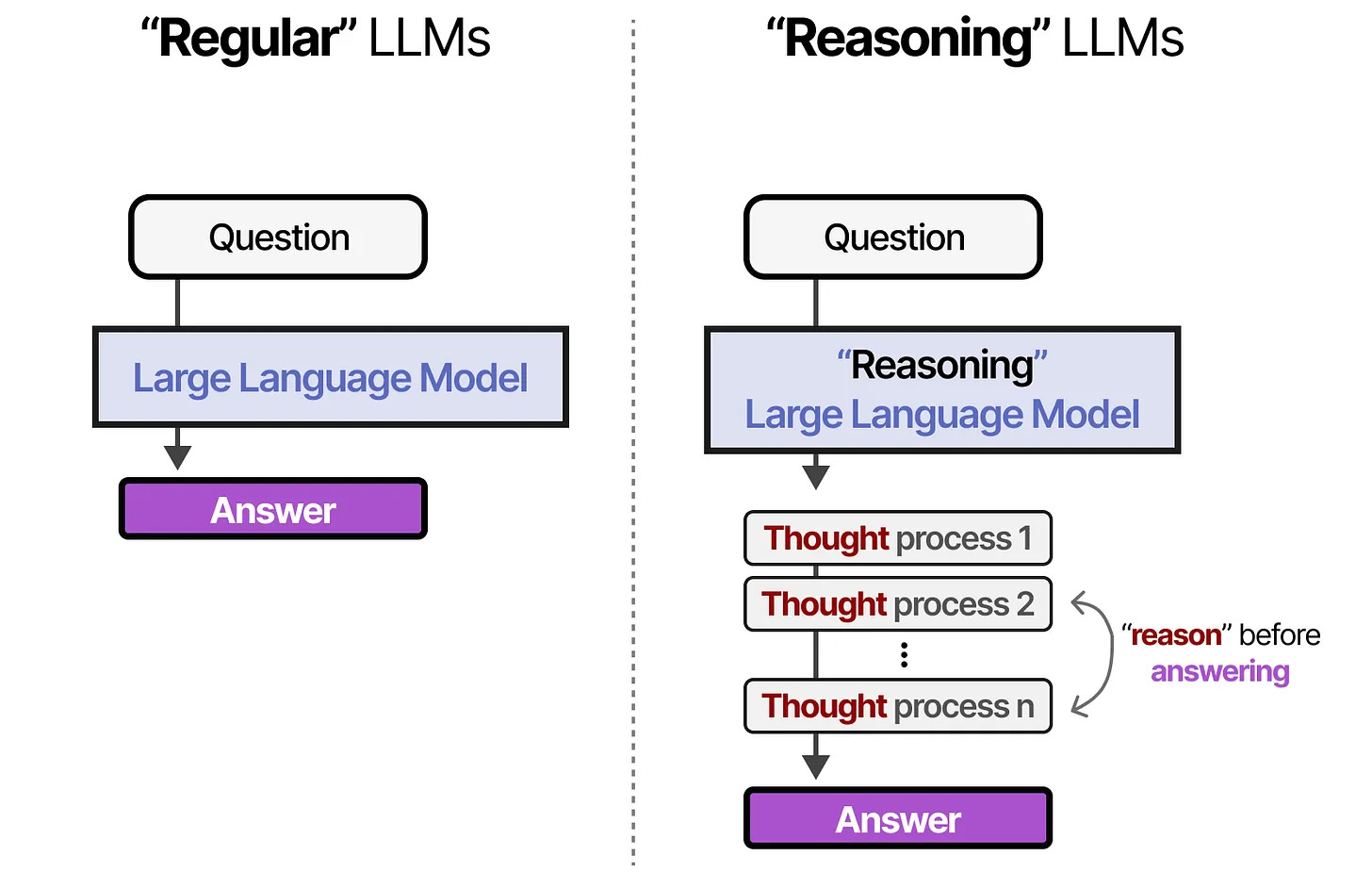}
    \caption{CoT Reasoning LLM vs Regular LLM~\cite{grootendorst2025visualguide}}
    \label{fig:cot}
\end{figure}

\subsection{Chain-of-Thought Reasoning LLMs} 

As shown in \Cref{fig:cot}, Chain-of-Thought (CoT) Reasoning LLMs~\cite{grootendorst2025visualguide} differ from regular LLMs by breaking down problems into smaller steps or thought processes before generating responses. This structured approach to inference has already been applied to user modeling through Chain-of-thought prompting for personalization~\cite{yang2024chain}. This method enables LLMs to reason about user attributes, subjective preferences, and intentions to create comprehensive user models expressed in natural language~\cite{li2024identify}. Although recent advances such as DeepSeek R1~\cite{guo2025deepseek} have demonstrated the inherent capacity of LLMs for chain-of-thought reasoning, the application of CoT reasoning LLMs  specifically for personalization and user modeling remains relatively unexplored, presenting an opportunity for further investigation.

\begin{table*}[htbp]
\footnotesize
    \centering
     \caption{Example of Data}
    \label{tab:my_label}
    \begin{tabular}{p{1cm}|p{4.5cm}|p{4.5cm}|p{6cm} }
    \toprule
   \textbf{Image} & \textbf{User Profile} & \textbf{Question} & \textbf{Answer} \\ \midrule
      \raisebox{-\totalheight}{\includegraphics[width=1cm]{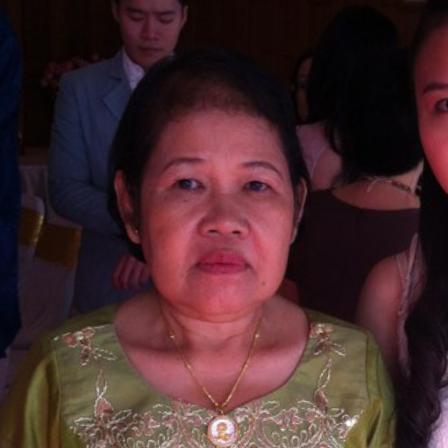}} & The person appears to be a southeast Asian female, approximately 60 to 69 years old. & Are there assistance services for people with mobility difficulties? &  Yes, many countries offer mobility assistance services, including specialized transport and home support, tailored to seniors' needs. \\ \midrule
    
\raisebox{-\totalheight}{\includegraphics[width=1cm]{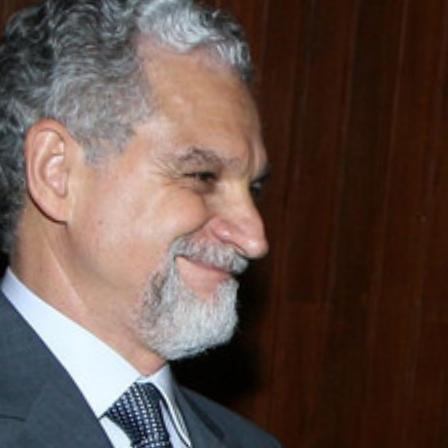}} & The person appears to be an Indian male, approximately 60 to 69 years old. & How do I report a fraudulent email? &
You can report a fraudulent email by forwarding it to your email provider's abuse department or using the "Report as Spam" feature in your email client.  \\ \bottomrule
    \end{tabular}
   
\end{table*}
\section{METHODS}
  Our framework employs a RAG architecture to create personalized user experiences through multimodal understanding. The system processes facial images and queries to produce contextually relevant, personalized responses. \Cref{fig:arch} illustrates the architectural components and their interactions.

\subsection{User Encoder}
The User Encoder is designed to encode facial images of individuals and query the database to identify and retrieve the corresponding user model and prior conversational context, facilitating seamless integration into subsequent dialogue prompts for enhanced interaction.
The User Encoder is a multimodal pre-trained model designed to encode user profiles by extracting visual features from facial images. It employs a vision encoder function $E_I: \mathbb{R}^{d_I \times N} \rightarrow \mathbb{R}^{d_h \times N}$, based on a Transformer architecture, to process images represented as sequences of tokens in feature vectors. A special \texttt{cls} token is used to derive a global image representation, which is subsequently mapped to an embedding space through a multilayer perceptron projection head $P: \mathbb{R}^{d_h} \rightarrow \mathbb{R}^{d_e}$, which yields the embedding $e_I$. Cosine similarity is utilized to measure the relationships between embeddings, and the model is trained using a contrastive loss function $\mathcal{L}_{\text{c}}$.

\subsection{User-VLM}
The User-VLM~\cite{rahimi2025user} initiates user modeling in cold-start scenarios where a user model is unavailable. By leveraging facial images and user queries, it generates an initial user profile comprising attributes such as age, gender, emotion, and ethnicity. These models employ a Llava-style architecture that seamlessly combines a vision encoder with a large language model. The model undergoes user-aware fine-tuning to enhance its ability to understand user-specific contexts within visual-linguistic interactions. This user-aware tuning enables the model to deliver tailored responses that align with individual preferences and visual content. Formally, the vision encoder $\mathcal{E}$ processes user images $X_I$ into vision representations $\mathbf{H}_I \in \mathbb{R}^{d_I}$, while the LLM generates textual user profile outputs $\mathbf{y} = \{y_1, y_2, \ldots, y_L\}$ based on tokenized prompt $\mathbf{H}_Q \in \mathbb{R}^{d_Q}$ and image embeddings $\mathbf{H}_I$. 

\subsection{Chain-of-Thought Reasoning LLM}
The CoT Reasoning LLM performs two critical functions: dynamically updating user profiles during interaction and providing personalized responses. This component can be defined as $F: \mathcal{X} \to \mathcal{S} \times \mathcal{Y}$. Given an input $x \in \mathcal{X}$, it generates an ordered sequence of intermediate reasoning steps $s = (s_1, s_2, \dots, s_k)$ along with a final answer $y \in \mathcal{Y}$. Under a probabilistic framework, this process is modeled by the joint distribution:

\begin{equation}
P(s, y \mid x) = \prod_{i=1}^{k+1} P(z_i \mid x, z_1, \dots, z_{i-1})
\end{equation}
where for $1 \leq i \leq k$, $z_i = s_i$ (the reasoning steps) and $z_{k+1} = y$ (the final answer). This formulation enables the model to not only provide appropriate responses but also maintain a dynamic understanding of user characteristics and preferences through explicit reasoning. The reasoning LLM takes as input the initial user profile generated by the User-VLM, reassesses this profile based on ongoing interactions, and produces contextually appropriate responses tailored to the updated user understanding.

\section{EXPERIMENTS}

\subsection{Dataset}
\label{sec:dataset}
To evaluate the personalization capabilities of the proposed model, we utilize \textit{ElderlyTech-VQA Bench}~\cite{rahimi2025user}, which comprises 144 triplets of images, questions, and answers, focusing on real-world questions posed by elderly individuals about technology. According to \cite{rahimi2025user}, the associated images, sourced from the FairFace dataset, are curated to ensure diversity in ethnicity and gender representation. Additionally, reference answers for user queries were generated using GPT-4o, following detailed instructions to deliver high-quality and contextually relevant responses. Examples of this dataset is shown in \Cref{tab:my_label}.

\begin{table*}[htbp]
\caption{\textbf{Comparative Study Results:} Our framework achieves significantly superior performance in ROUGE F1 scores comparing to baseline, with balanced precision and recall.}
\centering
\begin{tabular}{c|ccc|ccc|ccc}
\toprule
\multirow{2}{*}{\textbf{Model}} & \multicolumn{3}{c}{\textbf{ROUGE-1}} & \multicolumn{3}{c}{\textbf{ROUGE-2}} & \multicolumn{3}{c}{\textbf{ROUGE-L}} \\ \cmidrule(r){2-4} \cmidrule(r){5-7} \cmidrule(r){8-10} 
                                & Precision    & Recall    & F1        & Precision    & Recall    & F1        & Precision    & Recall    & F1        \\ \midrule
LLaMA 3.2                       & 0.1420       & 0.6068    & 0.2211    & 0.0514       & 0.2344    & 0.0817    & 0.1075       & 0.4629    & 0.1676    \\
Pixtral                         & 0.1489       & 0.6030    & 0.1934    & 0.0453       & 0.2270    & 0.0641    & 0.1117       & 0.4590    & 0.1448    \\
LLaVA-v1.6                      & 0.0956       & 0.6958    & 0.1658    & 0.0355       & 0.2685    & 0.0620    & 0.0730       & 0.5346    & 0.1267    \\
LLaVA-v1.5                      & 0.1256       & 0.6302    & 0.2035    & 0.0410       & 0.2207    & 0.0675    & 0.0943       & 0.4817    & 0.1535    \\ \midrule
\textbf{User-LLM R1 (ours)}                & \textbf{0.4294}       & \textbf{0.5167}    & \textbf{0.4531}    & \textbf{0.1424}       & \textbf{0.1677}    & \textbf{0.1485}    & \textbf{0.2376}       & \textbf{0.2799}    & \textbf{0.2478}   \\ \bottomrule
\end{tabular}
\label{tab:comp}
\end{table*}

\begin{table*}[htbp]
\caption{\textbf{Ablation Study Results:} while the size of the User-VLM model (3B vs. 10B) has minimal impact on F1 scores, larger reasoning models significantly improve performance, with higher parameter counts yielding superior ROUGE metrics.}
\centering
\begin{tabular}{cc|ccc|ccc|ccc}
\toprule
\multirow{2}{*}{Reasoning Model} & \multirow{2}{*}{User-VLM Model} & \multicolumn{3}{c}{ROUGE-1} & \multicolumn{3}{c}{ROUGE-2} & \multicolumn{3}{c}{ROUGE-L} \\ \cmidrule(r){3-5} \cmidrule(r){6-8} \cmidrule(r){9-11}
                                               &                                 & Precision & Recall & F1     & Precision & Recall & F1     & Precision & Recall & F1     \\ \midrule
\multirow{2}{*}{Distill-Qwen-1.5B}             & 3B                              & 0.4058    & 0.2809 & 0.3141 & 0.0876    & 0.0592 & 0.0665 & 0.2262    & 0.1498 & 0.1700 \\
                                               & 10B                             & 0.4122    & 0.2702 & 0.3064 & 0.0922    & 0.0599 & 0.0679 & 0.2389    & 0.1475 & 0.1703 \\ \midrule
\multirow{2}{*}{Distill-Qwen-7B}               & 3B                              & 0.4323    & 0.3768 & 0.3854 & 0.1130    & 0.0977 & 0.1005 & 0.2391    & 0.2016 & 0.2084 \\
                                               & 10B                             & 0.4319    & 0.3895 & 0.3971 & 0.1097    & 0.0972 & 0.0996 & 0.2301    & 0.2029 & 0.2088 \\ \midrule
\multirow{2}{*}{Distill-Llama-8B}              & 3B                              & 0.4851    & 0.3709 & 0.4056 & 0.1520    & 0.1140 & 0.1255 & 0.2743    & 0.2046 & 0.2264 \\
                                               & 10B                             & 0.4931    & 0.3837 & 0.4196 & 0.1539    & 0.1149 & 0.1275 & 0.2776    & 0.2092 & 0.2319 \\ \midrule
\multirow{2}{*}{Distill-Qwen-14B}              & 3B                              & 0.5035    & 0.4315 & 0.4498 & 0.1688    & 0.1428 & 0.1495 & 0.2913    & 0.2449 & 0.2576 \\
                                               & 10B                             & 0.4922    & 0.4238 & 0.4392 & 0.1641    & 0.1385 & 0.1448 & 0.2851    & 0.2378 & 0.2501 \\ \midrule
\multirow{2}{*}{Distill-Qwen-32B}              & 3B                              & 0.4906    & 0.4471 & 0.4541 & 0.1612    & 0.1449 & 0.1477 & 0.2801    & 0.2490 & 0.2557 \\
                                               & 10B                             & 0.4828    & 0.4396 & 0.4505 & 0.1580    & 0.1428 & 0.1468 & 0.2721    & 0.2444 & 0.2518 \\ \bottomrule
\end{tabular}
\label{fig:abl}
\end{table*}

\subsection{Metrics}
We selectively employ ROUGE~\cite{lin-2004-rouge} metrics to evaluate the framework across different types of questions, as their use provides a robust assessment of both factual consistency (via lexical overlap) ensuring outputs meet the dual demands of accuracy and adaptability in human-robot collaboration. We also compare the ground truth responses with the proposed model through human expert evaluation.

\subsection{Baseline}
The proposed model is evaluated against four state-of-the-art models of comparable size to ensure a rigorous and fair comparison. The first model, LLaMA 3.2 Vision~\cite{dubey2024llama}, is an advanced architecture based on CLIP~\cite{radford2021learning} and LLaMA 3.1, comprising 11 billion parameters. The second model, Pixtral~\cite{agrawal2024pixtral}, features a 12-billion-parameter multimodal decoder built upon Mistral NeMo~\cite{mistral_nemo}, along with a 400-million-parameter vision encoder trained from scratch. Additionally, the third and fourth models, LLaVA 1.5~\cite{liu2024visual} and LLaVA 1.6~\cite{liu2024improved}, employ Mistral~\cite{jiang2023mistral} and Vicuna~\cite{touvron2023llama} as their respective backbones, each comprising 7 billion parameters and integrating a CLIP-based vision encoder.

\subsection{Experimental Setting}
In our experiment, we employ six dense models distilled from DeepSeek-R1~\cite{guo2025deepseek}, leveraging both Llama and Qwen architectures, with model sizes ranging from 1.5B to 70B parameters. All experiments were conducted on a MacBook Pro M4 MAX with 64 GB of unified memory. During evaluation, we used greedy decoding with a temperature of 1.0 to generate responses. The maximum sequence length was set to 512 tokens for both input and output.

\subsection{Human Evaluation Methodology}

The human evaluation involved three experts assessing answers from two models—GPT-4o (the ground truth of the dataset) and the proposed framework based on Llama 3.3 70B R1—on 144 questions from the \textit{ElderlyTech-VQA Bench}. These questions, asked by elderly individuals, were tailored to specific profiles (age, gender, and ethnicity). Two evaluators were domain-specific experts with experience in elderly-accessible technologies, while the third was a general technology expert. All evaluators were native French speakers, and the English responses were translated into French, with possible translation biases considered during scoring. The questions addressed the use of digital and technological services, focusing on challenges elderly individuals might face, such as scanning documents or filing taxes online. Responses were rated on a scale of 1 (not relevant) to 5 (personalized), based on relevance, adaptability to the questioner's profile, and suitability for an elderly audience. Special attention was given to age-appropriate complexity, avoiding stereotypes in personalization for gender and ethnicity. The evaluation was conducted blindly, with evaluators unaware of which model provided each response. Each expert assessed 288 answers (two responses per question), totaling 864 scores across all evaluators.


\section{RESULTS}

As presented in \Cref{tab:comp}, the proposed framework based on \textit{Distill-Llama 3.3 70B R1} demonstrates significant improvements over the baseline models across three variations of the ROUGE metric. Specifically, the model achieves higher F1 scores, with notable gains in ROUGE-1 (0.4531) and ROUGE-L (0.2478), showcasing its enhanced capability in capturing relevant information. In contrast, baseline models such as LLaMA 3.2 and Pixtral exhibit comparatively lower scores across all metrics, indicating the superior performance and robustness of the proposed approach in text summarization tasks. The high recall but low precision in other models often indicates that they are prone to over-generation, meaning they tend to label or predict more positives than necessary. In simpler terms, these models may predict positive almost all the time, leading to high recall (capturing all true positives) but at the expense of precision (many false positives are included). The balanced precision and recall in our result is a positive outcome as it suggests that the proposed framework neither over-predicting nor under-predicting the positive class excessively.

As depicted in \Cref{fig:abl}, the ablation study examines 12 experimental variations, involving six reasoning models of varying parameter sizes and two configurations of the User-VLM model (3B and 10B). The results indicate that the evaluation scores, particularly F1, are not significantly influenced by the size of the User-VLM model, as comparable performances are observed between the 3B and 10B configurations. However, the number of parameters in the reasoning model has a direct and notable impact on performance, with larger models consistently achieving higher F1 scores across all ROUGE metrics. For instance, Distill-Qwen-14B and Distill-Qwen-32B outperform smaller models like Distill-Qwen-1.5B, underscoring the benefits of increased parameterization in reasoning tasks.

\begin{figure}
    \centering
    \includegraphics[width=1\linewidth]{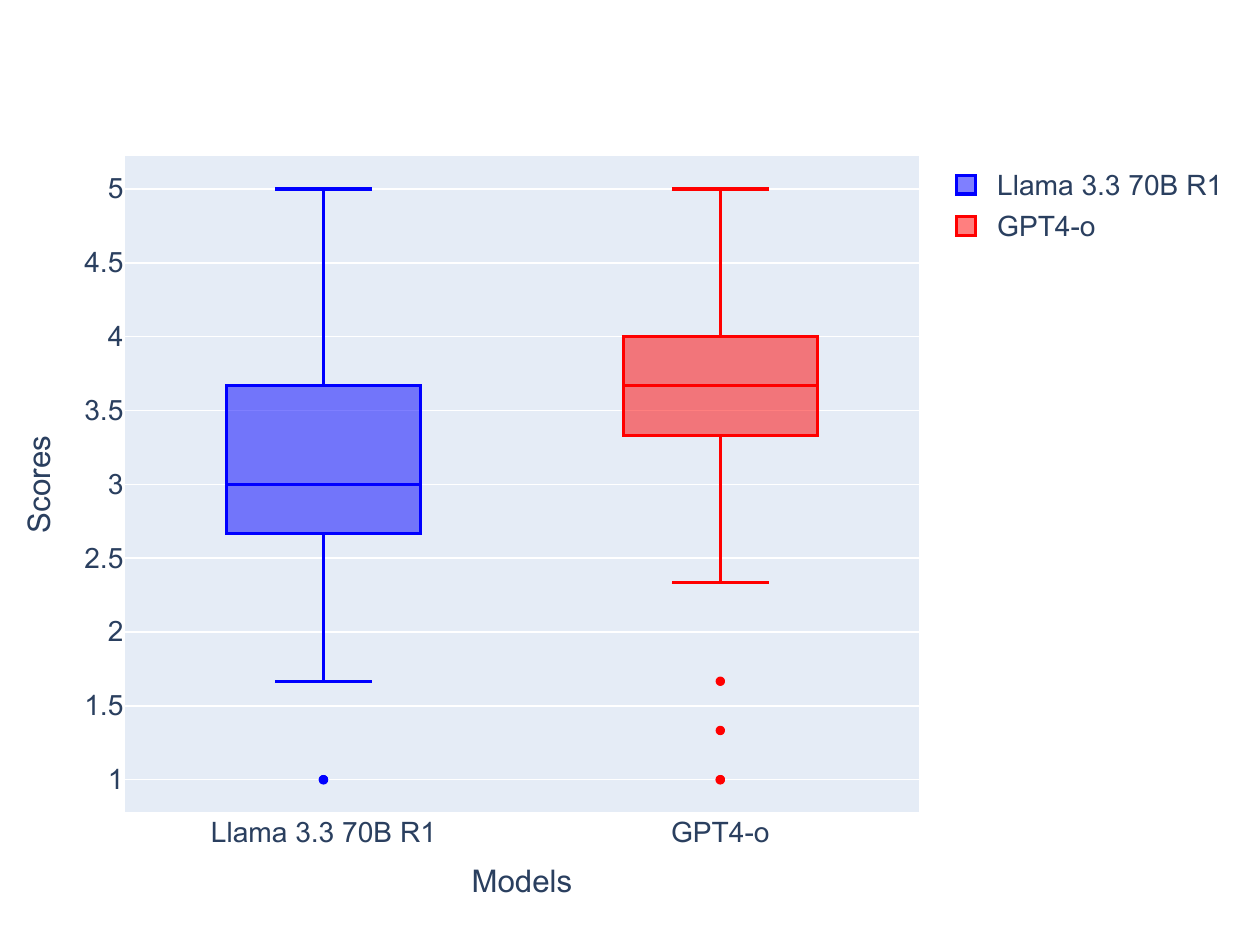}
    \caption{\textbf{Evaluation with Human Expert:} Our framework with 70B parameters and an adaptive User-VLM module, achieves GPT-4o-level personalization at lower computational cost.}
    \label{fig:human_eval}
\end{figure}

As illustrated in \Cref{fig:human_eval}, the proposed framework with 70B active parameters achieves performance comparable to GPT-4o, which utilizes approximately 200B parameters \cite{abacha2024medec}, particularly in terms of personalization. A key distinction between the two models lies in how user profiles are incorporated. In GPT-4o, the user profile is directly provided to the model as part of the dataset, simplifying the task of generating personalized responses. Conversely, our framework leverages the innovative User-VLM module to dynamically extract user profiles. This approach not only enables adaptive personalization but also offers a scalable method for accommodating new user profiles without requiring pre-defined data. The reliance on User-VLM ensures that the personalization process is more robust and context-aware, which is critical for applications targeting elderly individuals who may have varying levels of technological literacy, cultural backgrounds, and personalized needs. Moreover, the significant reduction in parameter count—70B compared to 200B—highlights the efficiency of the proposed framework, suggesting that it can achieve comparable levels of personalization while being computationally less demanding.

\section{ETHICAL CONSIDERATIONS}
Generally speaking, VLMs that analyze facial characteristics raise significant ethical concerns at both individual and societal levels. A primary issue is the inference of cultural preferences, personality traits, and social behaviors from physical characteristics. If datasets are biased or not representative, they can lead to misinterpretation, thus resulting in reductionist or offensive categorizations that do not represent individuals. On a larger scale, such practices can reinforce systemic discrimination by associating physical features with social stereotypes~\cite{rhue_racial_2018,benjamin_race_2020}. Furthermore, the collection of biometric data presents serious privacy and consent issues. Unless individuals are adequately informed and/or explicit consent is given, their basic rights, as protected by the GDPR~\cite{regulation2019gdpr} and the European Convention on Human Rights (ECHR)~\cite{harris2023law} for instance, can be violated. Moreover, the danger of exploitation of such data in surveillance or social control mechanisms is another critical issue~\cite{zuboff_age_2019,introna_picturing_2004}. Potential abuses can also include the threat of repression and criminalization by facial recognition ~\cite{bowyer_criminality_2020}, the discriminatory profiling of migrants or travelers at border controls~\cite{molnar_new_2019}, or the potential exacerbation for discrimination in recruitment processes~\cite{chen_ethics_2023}.

The framework presented here allows for the mitigation or avoidance of some ethical risks. The CoT Reasoning LLM provides constant updates of the user profile and personalized responses according to the interaction. It allows for a dynamic user profiling through the collection of user feedback, and thereby avoids excessive stereotyping based solely on physical features. Even during the interaction between the user and the LLM, the model is trained to offer several responses including some that are not related to facial features. Furthermore, explicit consent is always obtained before any data analysis. 

However, ethical challenges still exist. Algorithmic bias due to historical data can never be fully eradicated, and it is difficult to precisely know if meaningful consent is really given~\cite{dubber_oxford_2020}. Continuous monitoring and the reassessment of data are essential to promote fairness and equity~\cite{mehrabi_survey_2021}. Additionally, the use of these technologies must be regulated through strict legal frameworks to prevent abuse~\cite{european_data_protection_supervisor_annual_2021}.

\section{CONCLUSIONS}
This paper introduces USER-LLM R1, a novel framework designed to tackle the cold start problem in personalized human-robot interactions. By integrating chain-of-thought reasoning with vision-language models, the system dynamically generates and refines user profiles, enabling contextually relevant and adaptive responses from the initial encounter. Our evaluations on the \textit{ElderlyTech-VQA Bench} demonstrate significant improvements in ROUGE metrics over state-of-the-art baselines, highlighting the efficacy of our multi-stage reasoning approach. Ablation studies further underscore the importance of reasoning model size in achieving enhanced performance. Human evaluations validate the framework’s ability to generate personalized responses comparable to GPT-4o, particularly for elderly users, fostering engagement and trust. While ethical considerations regarding privacy and bias are thoroughly addressed, continuous monitoring and refinement are essential for responsible deployment. Future work will focus on expanding the framework to incorporate additional modalities and enhancing its adaptability to diverse user populations, ensuring equitable and effective human-robot interactions.



\section*{ACKNOWLEDGMENT}
The authors sincerely acknowledge the financial support of the French National Research Agency (ANR) for the ANITA project (Grant No. ANR-22-CE38-0012-01). 
\bibliographystyle{ieeetr}

\bibliography{ref}
\end{document}